\input{aipcheck}
\documentclass[
    ,final            
  ]
  {aipproc}

\layoutstyle{6x9}

\usepackage{color}

\begin{document}
\title{Cosmology today--A brief review}
\classification{98.80.-k;98.80.Bp;98.80.Es}
\keywords{Theoretical cosmology, Observational cosmology}

\author{Jorge L. Cervantes-Cota}{address={Depto. de F\'{\i}sica, Inst. Nac. de Investigaciones
Nucleares, A.P. 18-1027,  11801 M\'{e}xico DF, \\ Berkeley Center for Cosmological Physics, University of California, 
Berkeley, USA.}}

\author{George Smoot}{address={Lawrence Berkeley National Laboratory and University of 
California at Berkeley, Berkeley, USA, \\
Institute for the Early Universe WCU at Ewha Womans University Seoul, Korea,\\
Paris Centre for Cosmological Physics, Universite Paris Diderot, France.}}

\begin{abstract}
This is a brief review of the standard model of cosmology. We first introduce the FRW models and their flat 
solutions for energy fluids playing an important  role in the dynamics at different epochs. We then introduce different 
cosmological lengths and some of their applications. The later part is dedicated to the physical processes 
and concepts  necessary to understand the early and very early Universe and observations of it.
\end{abstract}

\maketitle


\section{Introduction \label{intro}}

The purpose of the present review is to provide the reader with an  
outline of modern cosmology. This science is passing through 
a revolutionary era, mainly because recent high precision
observations have severely constrained theoretical speculations, and
opened new windows to the cosmos. Tracking the recent history, in the late 1940's 
George Gamow \cite{Ga46}  predicted that the Universe should had 
begun from a very dense state, characterized by a huge density at
very high temperatures, a scenario dubbed the {\it Big Bang}, that was conjectured by George Lema\^{i}tre 
in the early 30's.  This scenario predicts that matter and light were at very high energetic
states, and both components behaved as a radiation fluid, in thermal equilibrium 
described as Planck blackbody (Bose-Einstein and the related Fermi-Dirac distributions).
This initial state remains today  imprinted
in the Cosmic Microwave Background Radiation (CMBR). 
Gamow's scenario predicted  this primeval radiation would be measured at a temperature
of only few Kelvin's degrees; since the expansion of the
Universe cools down any density component. Robert Dicke, and others,
begun the race to discover this radiation coming from the cosmos,
and in 1965 A. A. Penzias and R.W. Wilson of the Bell telephone
laboratories discovered, by chance, this radiation form, confirming
the general Big Bang scenario; see the original references published
in Ref. \cite{KoTu88}.

During this elapsed period the Big Bang scenario was generally accepted.
However, some key questions remained open, for instance, whether or not
this radiation was of Planckian nature to entirely confirm  that the
Universe was in thermal equilibrium at the very beginning of time. 
In 1990 a modern version of the Penzias--Wilson experiment was carried out.  
This experiment, lead by G. Smoot and J. Mather with the
Cosmic Background Explorer (COBE) satellite, started a new {\it high
precision} experimental era in cosmology \cite{Mather90}. 
The COBE team for the first time revealed that the Universe was in equilibrium 
(the radiation had the Planck form to high precision), and was almost
homogeneous and isotropic, but not completely\cite{Sm92}.  The tiny ($10^{-5}$) 
anisotropies found by COBE  --also imprinted in the matter distribution-- 
were lately the responsible for the formation of stars, galaxies, clusters, and 
all large scale structures of our Universe.

The origin of these tiny anisotropies is presumably in quantum
fluctuations of fundamental fields of nature which were present in
the very early Universe.  Modern quantum field theories, together
with cosmological models, help to understand how this small
fluctuations evolved to become of cosmic scales, enabling COBE to
detect them.  This satellite and other more recent cosmological
probes, such as BOOMERANG, MAXIMA, WMAP and now PLANCK satellite that measured CMBR  and  the
Two degree Field (2dF) galaxy survey and the Sloan Digital Sky Survey (SDSS), among others, 
not only confirmed with a great accuracy some of the theoretical predictions of the standard
big bang cosmological model, but also opened the possibility to test
theories and scenarios applicable in very early Universe, such as 
inflation, or in present times, such as quintessence. In this way,
cosmology that used to be a purely theoretical and often speculative science, is today
subject to high precision tests in light of these new 
observations \cite{BrCeSa04,AmTs10}.

\section{On the standard model of cosmology \label{sc} }

We  begin our study by reviewing some aspects of the standard 
lore of physical and theoretical cosmology.  In doing that we 
consider the Friedmann-Robertson-Walker (FRW)  model 
in Einstein's general relativity (GR) theory.  We shall make use of 
``natural'' units $\hbar = c = k_{B} = 1$ and our geometrical sign conventions are 
as in Ref. \cite{MiThWh73}. 

\subsection{FRW models} 
The {\it cosmological principle} states that the Universe is both spatially homogeneous and isotropic on the large scale,
which was originally assumed but now observed to be valid for the very large large scale of the Universe.
This homogeneous 
and isotropic space--time symmetry was originally studied by Friedmann, 
Robertson, and Walker (FRW), see Refs. \cite{Fr22Ro35Wa36}.  The symmetry is 
encoded in and defines the unique form of the line element:

\begin{equation}						
 ds^{2} = -dt^{2} + a^{2}(t) \,\  \left[ \frac{dr^{2}}{1-k r^{2}}  + r^{2} (d\theta^{2} + {\rm sin}^{2} \theta \, d\phi^{2}) \right] , 
 \label{frwmsc}	\end{equation}					
where $t$ is the cosmic time, $r-\theta-\phi$ are polar coordinates, which can be adjusted so that the constant curvature 
takes the values $k$ = 0, +1, or $-1$ for a flat, closed, or open space, respectively. $a(t)$ is the unknown potential of the 
metric that encodes the size at large scales, more formally is the {\it scale factor} of the Universe.

To then solve this line element for the scale factor of the Universe $a(t)$ and perturbations one simply needs a valid theory of gravity.
The standard model of cosmology is based on GR, which can be 
derived from the Einstein-Hilbert Lagrangian
\begin{equation}	
{\cal L} = \frac{1}{16 \pi G} (R + L_{m}) \sqrt{-g} \,\ ,
\label{elsc}  \end{equation} 
where $R$  is the Ricci scalar, $G$ the Newton constant, and $g=|g_{\mu \nu}|$ 
the determinant of the metric tensor.   By performing the metric variation of this equation, one obtains the 
well known Einstein's field equations 
\begin{equation}						
 R_{\mu \nu} -\frac{1}{2}R g_{\mu \nu} =  8\pi G T_{\mu \nu} \,\ ,
 \label{eesc}	\end{equation}					
where $T_{\mu \nu} \equiv - \frac{2}{\sqrt{-g}} \frac{\partial L_{m} \sqrt{-g}}{\partial g^{\mu \nu}}$ is the stress  
energy--momentum tensor, which is a symmetric tensor.  Thus,  Eq. (\ref{eesc}) is a collection of ten coupled partial 
differential equations. However, the theory is diffeomorphism invariant, and one adds to them a 
gauge condition, implying in general four extra equations to Eq. (\ref{eesc}) that reduce the degrees of freedom.



The beautifully symmetric FRW solutions to the Einstein Eqs. (\ref{eesc}) represent a cornerstone in 
the development of modern cosmology, since with them it is possible to 
understand the expansion  of Universe as was realized in the 1920's through 
Hubble's law of expansion \cite{Hu27}.  With this FRW metric, the GR cosmological
field equations are,
\begin{equation} 	
H^{2} \equiv \left(\frac{\dot{a}}{a}\right)^{\! 2}= \frac{8 \pi G}{3} \rho - 
\frac{k}{a^{2}} 
 \label{frw1sc} \end{equation} 
and 
\begin{equation}
\frac{\ddot{a}}{a} = - \frac{4 \pi G}{3} (\rho + 3 p) \,\ , 
\label{frw2sc} \end{equation}  	
where $H$ is the {\it Hubble parameter}; $\rho $ and $p$ are the density and 
pressure of the perfect fluid considered, that is,  
$T_{\mu\nu}  =  \rho u_{\mu} u_{\nu} + p ( u_{\mu} u_{\nu} + g_{\mu \nu} )$, 
where $u_{\mu}= \delta^{0}_{\mu}$ is the fluid's four velocity in comoving 
coordinates.  Dots stand for cosmic time derivatives. 

Assuming energy--momentum tensor conservation, $T_{\mu \,\ ;\nu }^{~ \nu } = 0$, 
is valid, one obtains the continuity equation,
\begin{equation}
\dot{\rho} + 3 H (\rho + p) = 0 \,\ .
 \label{frw3asc}  \end{equation}

The above system of equations (\ref{frw1sc}), (\ref{frw2sc}), and (\ref{frw3asc}) implies three unknown  
variables ($a$, $\rho$, $p$) for three equations, but the equations are not all linearly independent,  just two 
of them. Thus, an extra assumption has to be made to close the system.  The answer should come from
the micro-physics of the fluids considered. For the moment let us assume a barotropic equation of state that is 
characteristic  for different cosmic fluids, 
$w = {\rm const.}$,
\begin{equation}
\frac{p}{\rho} = w = 
\left\{ \begin{array}{l@{\quad {\rm for} \quad}l}
\frac{1}{3}     &    {\rm radiation ~or ~relativistic ~matter}\\
   0            &    {\rm dust }\\
   1            &    {\rm stiff ~fluid }\\
 - 1            &    {\rm cosmological ~constant }  \end{array} \right.
\label{ecsc} \end{equation} 	
to integrate Eq. (\ref{frw3asc}),  yielding  
\begin{equation}
\rho = \frac{M_{w}}{a^{3(1+w )}}  ~~~~~~~{\rm or} ~~~~~~   \frac{\rho_x}{\rho_{x~0}} = \left( \frac{a_0}{a} \right)^{3(1+w )}  \,\ , 
 \label{frw3bsc} \end{equation} 	
where $M_{w}$ is the integration constant and is different dimensioned by
considering different $w-$fluids.  With this equation the system is closed and can be solved once the initial 
conditions are known.  

In addition to the above fluids, one can include an explicit {\it cosmological constant} ($\Lambda$) 
in  Eq. (\ref{frw1sc}) and arrange it in the following form:
\begin{equation} \label{Omega}
\Omega \equiv \Omega_R +  \Omega_M + \Omega_\Lambda  = 1 + \frac{k}{a^{2}
H^{2}}
\end{equation}
with $\Omega_R$ being the radiation component dominated by the CMB and small ($5 \times 10^{-5}$ at present), 
$\Omega_M \equiv \frac{8 \pi G \rho_m}{3 H^2}$ and
$\Omega_\Lambda \equiv \frac{\Lambda}{3 H^{2}}$. The
parameter $\Omega$ is called the  {\it density parameter},
which is composed of a matter part and a cosmological constant term.
Thus, we see that the different values of the density parameters
$(\Omega_{m}, \Omega_{\Lambda})$ will impose different values for
the curvature term. If $\Omega > 1$, it turns out $k$ is greater
than zero, signifying a Universe with a positive curvature (closed
Universe). If $\Omega < 1$, then $k < 0$, this corresponds to a
negative curvature (open Universe). A critical value is obtained
obviously when $\Omega = 1$, then the spatial curvature is null, $k=
0$. The value of the energy density for which 
$\Omega = [\rho + \Lambda/(8 \pi G)]/\rho_{c}  = 1$ holds is known as the {\it critical
density}, $\rho_c \equiv 3 H^{2}/8 \pi G$. The last term in  Eq. (\ref{Omega}) can be defined 
as $ \Omega_{k} = -\frac{k}{a^{2} H^{2}}$ and thus the Friedman equation becomes a  
constriction for the density parameters $\sum_i \Omega_i =1$, and this expression holds at 
any time.   It is useful to define an alternative measure of the expansion of 
the Universe through the {\it redshift} ($z$), $1+z \equiv a_0/a(t)$, where $a_{0}$ is the scale factor at present and 
is the set to unity by convention.   Today $z_0 =0$ and in the early Universe the redshift grows. 
In terms of the redshift the density parameters are, from  Eq. (\ref{frw3bsc}),   
$\Omega_i = \Omega_{i}^{(0)} (1+z)^{3(1+w)}$; quantities  with a subindex or superindex ``0'' are 
evaluated at the present time.

Let us very briefly recall which $w-$values are needed to describe the 
different epochs of the Universe's evolution.   The assumption that 
$w =1/3$ is valid for a ``fluid'' of radiation  and/or 
of ultra-relativistic matter ($T \gg m$, $m$ being its rest mass). This epoch is of  
importance at the beginning of the hot big bang theory, where the material content of the Universe 
consisted of photons, neutrinos, electrons, and other massive particles with 
very high kinetic energy.  After some Universe cooling, some massive particles
decayed and others survived (protons, neutrons, electrons) whose masses  
eventually dominated over the radiation components (photon, 
neutrinos) at the {\it equality} epoch ($\rho_{\rm rel} = \rho_{m}$) at  $z_{\rm eq} \sim 3200$ \cite{wmap7}.  From 
this epoch and until recent efolds of expansion ($z_{\rm DE} \sim 1/2$)  the  main matter component   
produced effectively no pressure on the expansion and, therefore, one can accept a 
model filled with dust, $w =0$, to be representative for the energy content of the  
Universe in the interval $3200 < z < 1/2$.    The dust equation of state is then representative of inert, cold {\it dark matter} (CDM). 
Dark matter (DM) does  not (significantly)  emit light and therefore it is dark. Another possibility is that dark matter interacts 
weakly and is  generically called WIMP (Weakly Interacting Massive Particle), being the neutralino the most popular 
WIMP candidate. Another popular dark matter candidate is the axion, a hypothetical particle postulated to 
explain the conservation of the CP symmetry in quantum chromodynamics (QCD).  Back to the Universe evolution, from 
$z \sim 1/2$ and until now the Universe happens to be accelerating with an equation of state   $w \approx -1$, 
due to some constant energy that induces a cosmological constant, $\Lambda =  8 \pi G \rho =$const.  The 
cosmological constant is the generic factor 
of an inflationary solution, see the $k =0$ solution below, Eq. (\ref{isc}).    The details of the expansion are still unknown and 
it is possible that the expansion is due to some new fundamental field  (e.g. quintessence) that induces an 
effective $\Lambda (t) \sim$const.  One calls (M. Turner dubbed it) {\it dark energy} (DE) to this new element. Dark energy does not 
emit light nor any other particle, as so far known, it simply behaves as a (transparent) media that gravitates with an effective 
negative pressure.  The physics behind  dark energy or even the cosmological constant is 
unclear since theories of grand unification (or theories of everything, including gravity) generically 
predict  a vacuum energy associated with fundamental fields, 
$<0|T_{\mu \nu}|0>= <\rho>  g_{\mu \nu}$, that turns out to be very large.  This can be seen by summing 
the zero-point energies of all normal modes of some field of mass $m$, to  
obtain $ <\rho>\approx M^{4}/(16 \pi^{2})$, where $M$ represents some cutoff 
in the integration, $M \gg m$.  Then, assuming GR is valid up to the Planck ($Pl$) 
scale, one should take $M\approx 1/\sqrt{8 \pi G}$, which gives  
$ <\rho>=10^{71}$ GeV$^{4}$.  This term plays the role of an effective  cosmological constant of 
$\Lambda = 8 \pi G  <\rho> \approx M_{Pl}^{2} \sim 10^{38}$ GeV$^2$ which must
be added to the  Einstein equations (\ref{eesc}) and yields
an inflationary solution Eq. (\ref{isc}).  However, since the cosmological
constant seems to dominate the Universe dynamics nowadays, one has that
\begin{equation}
\Lambda \approx 8 \pi G \rho_{0} = 3 H_{0}^{2} \sim 10^{-83} {\rm GeV}{}^{2}. 
\label{lsc} \end{equation}
which is very small compared with the above value derived on dimensional 
grounds.  Thus, the cosmological constraint and theoretical expectations 
are rather dissimilar, by about 121 orders of mag\-ni\-tude!  Even if one 
considers symmetries at lower energy scales the theoretical $\Lambda$ is indeed 
smaller, but never as small as the cosmological constraint: 
$\Lambda_{GUT}\sim 10^{21}$ GeV$^2$, $\Lambda_{SU(2)}\sim 10^{-29}$ GeV$^2$.
This problem has been reviewed since decades ago  \cite{We89,CaPrTu92} and remains 
open.   


The ordinary differential equations system described above needs a set of 
initial, or alternatively boundary, conditions to be integrated.  One has to 
assume a set of two initial values, say, 
$(\rho(t_{*}), \dot{a}(t_{*}) ) \equiv (\rho_{*}, \dot{a}_{*})$
at some (initial) time $t_{*}$, in order to determine its evolution. The full 
analysis of it can be found in many textbooks \cite{We72,MiThWh73}.  Here, 
in order to show some physical, early Universe consequences we assume $k =0$, 
justified as follows: From Eqs. (\ref{frw1sc}) and (\ref{frw3bsc}) one notes 
that the  expansion rate, given by the Hubble parameter, is
dominated by the density term as $a(t) \rightarrow 0$, since 
$\rho \sim 1/a^{3(1+w )} > k / a^{2}$ for $w > -1/3$, that is, 
the flat solution is very well fitted at the very beginning of times.  
Therefore, assuming $k =0$,  Eq. (\ref{frw1sc}) implies 
\begin{eqnarray}
a(t) & = & [6 \pi G M_{w} (1+w)^{2}]^{\frac{1}{3(1+w )}} 
(t-t_{*})^{\frac{2}{3(1+w )}} \nonumber \\[5pt]
 & = & \left\{ \begin{array}{l@{\quad {\rm for} \quad}l}
 (\frac{32}{3} \pi G M_{\frac{1}{3}})^{1/4}~(t-t_{*})^{1/2} & w=\frac{1}{3} ~ 
{\rm radiation}\\
 (6 \pi G M_{0} )^{1/3} ~~~ (t-t_{*})^{2/3}    & w=0 ~ {\rm dust}\\
 (24 \pi G M_{1} )^{1/6} ~~ (t-t_{*})^{1/3}     & w= 1 ~{\rm stiff ~fluid }\\
\end{array} \right. \label{frwsolsc}\end{eqnarray} 	
and 
\begin{equation}
a(t) =  a_{*} e^{H t}  \qquad {\rm for}\qquad w=-1 ~{\rm cosmological ~constant } 
\label{isc} \end{equation} 	
where the letters with a subindex ``$*$'' are integration constants, 
representing quantities evaluated at the beginning of times, $t=t_{*}$.   To 
obtain Eq. (\ref{isc}), the argument given right above to neglect $k$  
is not anymore valid, since here $\rho= {\rm const.}$; that is,
from the very beginning it must be warranted that 
$H^{2} \approx \frac{8 \pi G}{3} \rho_{*} > k/a^{2}_{*}$, otherwise 
$k$ cannot be ignored. Nevertheless if $\Lambda$ is present, it will eventually dominate 
over the other decaying components, this is the so called {\it cosmological no-hair theorem} \cite{ChMo94}.      
A general feature of all the above solutions is that they are expanding, at different Hubble rates, 
$H= \frac{2}{3(1+w)} \frac{1}{t}$ for Eqs. (\ref {frwsolsc}) and $H = $const. for Eq. (\ref {isc}).

From Eq. (\ref{frwsolsc}) one can immediately see that at 
$t=t_{*} , \,\ a_{*} = 0$ and from Eq. (\ref{frw3bsc}), 
$\rho_{*} = \infty$, that is, 
the solution has a singularity at that time, at the Universe's 
beginning; this initial cosmological singularity is precisely the big bang singularity.  As 
the Universe evolves the Hubble parameter goes as $H \sim 1/t$, i.e., 
the expansion rate decreases; whereas the matter-energy content acts as an 
expanding agent, cf. Eq. (\ref{frw1sc}), it  decelerates the expansion, 
however, asymptotically decreasing, cf. Eqs. (\ref{frw2sc}) and 
(\ref{frw3bsc}).  In that way, 
$H^{-1}$ represents an upper limit to the age of the Universe; for instance, 
$H^{-1}=2t$ for $w =1/3$ and $H^{-1}=3t/2$ for $w =0$, $t$ being the 
Universe's age.

The exponential expansion (\ref{isc}) possesses no singularity (at finite times), being  the 
Hubble parameter a constant.  A fundamental ingredient of this inflation  
is that the right hand side of  Eq. (\ref{frw2sc}) is positive, 
$\ddot{a}>0$, and this is performed when $\rho+3 p < 0$, that is, one 
does not have necessarily to impose the stronger condition $w=-1$, but it suffices that $w <-1/3$, in order to have a 
moderate inflationary solution; for example, 
$w =-2/3$ it implies $a=a_{*} t^{2}$, a mild power-law inflation. 

Since the scale factor evolves as a smooth function of time (most of the time!), one is 
able to use it as a variable, instead of time, in such as a way that $d/dt = a\,  H \, d/da$.  This 
change of variable helps to integrate the continuity equation for non-constant $w(a)$ to obtain:
\begin{equation} \label{cont-sol}
\rho(a) = \rho_0 e^{-3 \int [1+w(a)] da/a} . 
\end{equation}
If, for instance, one parametrizes dark energy through an analytic function of the scale factor, $w(a)$, one immediately
obtains its solution in terms of
\begin{equation} \label{time-scale-factor}
t = \int \frac{1}{\sqrt{8 \pi G\rho(a)/3}}\frac{da}{a} .
\end{equation}
 
In cosmology, typical times and distances are determined mainly by the Hubble parameter, and in practice measurements 
are often related to redshift, as measured from stars, gas, etc.    It is then useful to express the 
Friedmann Eq. (\ref{frw1sc}) in terms of the redshift.  The standard model of cosmology considers a Universe filled 
baryons, photons, neutrinos, CDM, and a cosmological constant ($\Lambda$), and is termed  
$\Lambda$CDM for short.  For this model one obtains: 
\begin{equation} \label{hz1}
H^2 = H_{0}^2  \sum_i \Omega_{i}^{(0)} (1+z)^{3(1+w_i)} , 
\end{equation}
where $w_i$ is the equation of state parameter for each of the fluids considered.  The present contribution of the 
main energy components can be fitted from different cosmological probes, obtaining    
$ \Omega_{b}^{(0)} = 0.046\pm 0.002$,   $ \Omega_{\rm DM}^{(0)}  = 0.23\pm 0.01$, and 
$ \Omega_{\Lambda}^{(0)} = 0.73\pm 0.02$, together with a  Hubble constant of $70.4^{+1.3}_{-1.4} $ 
km/s/Mpc \cite{wmap7}.  Around $96\%$ of mater-energy the Universe is composed of dark components!

In general, if dark energy is a function  of the redshift, from Eq. (\ref{cont-sol}) one can generalize the above equation to: 
\begin{equation} \label{hz}
H(z)^2/H_{0}^2  =   \Omega_{m}^{(0)} (1+z)^{3} + \Omega_{\gamma}^{(0)} (1+z)^{4} +  
 \Omega_{k}^{(0)} (1+z)^{2} +  \Omega_{DE}^{(0)} {\sf f}(z)  \, , 
\end{equation}
where $m$ stands for dark matter and baryons, $\gamma$ for photons, and 
\begin{equation}
{\sf f}(z) = {\rm exp} \left[ 3 \int_{0}^{z} \frac{1+w(z')}{1+z'}   dz' \right] \, .
\end{equation}
Eq. (\ref{time-scale-factor})  gives the age of the  Universe in terms of the redshift, $H_0$, and the density parameters: 
\begin{equation}
t_0  = H_{0}^{-1}  \int_{0}^{\infty} \frac{dz}{(1+z) H(z)} .
\label{time_z}
\end{equation}
When combining different cosmological probes one obtains for the  $\Lambda$CDM model an age of 
$t_0  =13.75 \pm 0.11$ Gyr \cite{wmap7}.
\subsection{Cosmic distances and their measurements}
It is useful to write the FRW metric, Eq. (\ref{frwmsc}), in terms of a new distance coordinate ($\chi$)
\begin{eqnarray}
d s^2=-d t^2+a^2(t)\left[d \chi^2+f_k^2 (\chi) 
(d\theta^2+\sin^2\theta \, d\phi^2)\right]\,,
\label{frw_metric2}
\end{eqnarray}
where 
\begin{eqnarray}
f_k (\chi) =  \left\{\begin{array}{lll}
{\rm sin} \chi\,, \quad & k=+1\,, \\
\label{fk}
\chi\,, \quad & k=0\,, \\
{\rm sin h} \chi\,, \quad & k=-1\,.
\end{array} \right. 
\end{eqnarray}

Now we proceed to define some cosmic distances necessary to understand the cosmic physics. 

{\bf Causal horizon.}  The region of space that can be connected to some other region  
by causal physical processes, at most through the propagation of light, implies $ds^{2}=0$.  For 
the FRW Eq. (\ref{frwmsc}) or (\ref{frw_metric2}), in spherical coordinates  with 
$\theta , \phi =$const. implies that 
\cite{Ri56,We72}:
\begin{equation}
\chi_{H} =  \int^{\chi_H}_{0}  d\chi   = \int^{t_0}_{t} \frac{dt'}{a(t')}   =  \frac{1}{a_0 H_0} \int^{z}_{0} \frac{dz'}{E(z')}       
\label{co-dist} \end{equation}
this is the so-called {\it comoving distance}, being the distance between two points in the Universe 
in which the expansion is  factored out.  Now,  the {\it causal} or {\it particle horizon}, $d_{H}$ is given by:
\begin{equation}
d_{H}(t) = a(t) \, \int^{\chi_H}_{0}  d\chi =  a(t) \,  \int^{t}_{t_*} \frac{dt'}{a(t')}  .
\label{h1sc} \end{equation}

In order to analyze the whole horizon evolution, from nowadays ($t_{0}$) to
the Planck time ($t_{Pl}$), we have to consider all the Universe stages, but for brevity we shall not include 
the current accelerated expansion. We firstly compute the horizon for the matter
dominated era $t_{\rm eq.} \leq t \leq t_{0}$ and secondly for the radiation 
era $ t \leq t_{\rm eq.}$, because they are differently determined by 
Eq. (\ref{frwsolsc}), where we set $t_{*}=0$ for convenience.  For the matter
epoch one has $a(t)=a_{0}  (t/t_{0})^{2/3}$, then the first equation above gives
$\chi_{H}=\frac{3}{a_{0}} (t_{0}^{2}~ t)^{1/3}$; from the second equation one obtains
the horizon $d_{H}(t)=3 t=2 H^{-1}$.  For the radiation period, one has that 
$\chi_{H} =\frac{2}{a_{\rm eq.}} ( t_{\rm eq.} ~t)^{1/2}$ and 
$d_{H}(t)=2 t = H^{-1}$.  We see, for the matter dominated era, the causal 
horizon is twice the Hubble distance, $H^{-1}$ (sometimes called {\it Hubble horizon}), and 
they are equal to each  other during the radiation dominated era; therefore, one uses them 
interchangeably.  It is clearly seen for both eras that as 
$t \rightarrow 0$, the Universe is causally disconnected, being 
$a(t)>d_{H}(t)$.   But, on the other side, by that time the CMBR was already  highly 
isotropic and with a black body spectrum.  Then, one has to take for granted that the initial conditions for all small horizon volumes 
were very fine tuned to account for the present observed large angle CMBR levels of isotropy, with 
$\delta T/T \approx {\rm few} \times 10^{-5}$. This is the horizon problem.

{\bf  Event Horizon.} The {\it event horizon}, $d_{e}$, determines the region of space which will keep 
in causal contact after some time;  that is, it delimits the region from which one can 
ever receive (up to some time $t_{\rm max}$) information about events taking place 
now (at the time $t$):
\begin{equation}
d_{e}(t) = a(t) \int_{t}^{t_{\rm max}} \frac{dt'}{a(t')} \,\ .
\label{eh1in} \end{equation}
For a flat model during its matter dominated era ($a\sim t^{2/3}$), 
$d_{e} \rightarrow \infty$ as $t_{\rm max} \rightarrow \infty$.  

{\bf  Luminosity distance.} The {\it luminosity distance} is the distance measured using the energy 
flux (${\cal F}$) observed by a light source with absolute luminosity ($L_s$):
 \begin{equation}
 \label{lu-dist}
d_{L}^2 \equiv \frac{L_s}{4\pi {\cal F}}\,,
\end{equation}
where ${\cal F} = L_0/S$, being $L_0$ the observed  luminosity and $S = 4\pi (a_0 f_{k}(\chi))^{2}$  is the 
sphere area at $z=0$.  The luminosity distance becomes
\begin{equation}
d_{L}^2  = (a_0 f_{k}(\chi))^{2} \frac{L_s}{L_0} \, .
\end{equation}

If we express the energy emitted by a light pulse in a time interval $\Delta t_1$ as $\Delta E_1$, the absolute luminosity is 
given by $L_{s} = \Delta E_{1}/\Delta t_{1}$. Similarly we define $L_{0} = \Delta E_{0}/\Delta t_{0}$, where $\Delta E_{0}$ 
is the detected energy in a time $\Delta t_{0}$.  On the other hand, since the photon energy can be expressed in terms of its 
wavelenght ($\lambda$), one has that $\Delta E_{1}/\Delta E_{0} = \lambda_{0}/ \lambda_{1} = 1+ z$  and moreover
$c=1$, being constant implies that $\lambda_{1}/\Delta t_{1} = \lambda_{0}/\Delta t_{0} $,
from which we finally have that   
\begin{equation}
\frac{L_s}{L_0} = \frac{\Delta E_1}{\Delta E_0} \frac{\Delta t_{0} }{\Delta t_{1} } = (1+z)^{2} , 
\end{equation}
and the luminosity distance becomes
\begin{equation} \label{dl}
d_{L}  = a_0 f_{k}(\chi) (1+z) . 
\end{equation}
Since $f_{k}(\chi)$ depends on $\chi$ and this on the redshift, cf. Eq. (\ref{co-dist}), thus by measuring 
the luminosity distance, we can determine the expansion rate of the Universe.  For the $\Lambda$CDM
model one finds that  \cite{CoSaTs06}  
\begin{equation}
d_{L}  = \frac{(1+z)}{H_0} \int_{0}^{z}  \frac{dz'}{\sqrt{\sum_i \Omega_{i}^{0} (1+z')^{3(1+w)}}} .
\end{equation}
The luminosity distance becomes larger when the cosmological constant is present and this is what was found to
fit better the supernovae Ia.  

In practice, one uses the relationship of the apparent ($m$) and absolute ($M$) 
magnitude, related to the luminosity  measured at present and when emitted, respectively to have:
\begin{equation} \label{mag}
m-M = 5\,  {\rm Log}_{10} \left(\frac{d_L}{\rm Mpc} \right) + 25. 
\end{equation}

By adjusting best fit curves to their data, two different supernova groups \cite{Ri99-Pe99}  found a clear evidence for 
$\Lambda$ in the late 90's.  The presence of a cosmological constant makes the Universe not only expanding, but accelerating
and, in  addition,  its age is older, not conflicting with globular cluster ages.  With the course of the years, various supernova  
groups have been getting more confidence  that the data is compatible with the presence of dark energy, dark matter, and 
a high value of Hubble parameter.   One of the latest data release, the Union2 
compilation \cite{Am10}, reports that the flat concordance $\Lambda$CDM 
model remains an excellent fit to the data with the best fit constant equation of state parameter 
$w=-0.997^{+0.050}_{-0.054}$ for a flat Universe, and 
$w=-1.035^{+0.055}_{-0.059} $ with curvature. Also, they found that $\Omega_m = 0.270 \pm 0.021$ (including baryons and DM) 
for fixed $\Omega_k=0$. That is, $\Omega_{\Lambda} = 0.730 \pm 0.021$.   

{\bf Angular diameter distance.} The {\it angular diameter distance}  is given by
\begin{equation}
d_{A} \equiv \frac{\Delta x}{\Delta \theta}, 
\end{equation}
where $\Delta \theta$ is the angular aperture of an object of size $\Delta x$ orthogonal to the line of 
sight in the sky.  Usually this distance is used in the CMBR anisotropy observations, since the source emitting  
the radiation is on a surface of a sphere of radius $\chi$ with the observer located at the center; it is also used in the 
determination of the BAO feature, see below.  Thus, the size  $\Delta x$ at the time $t_1$ in the Friedmann metric, Eq. (\ref{frw_metric2}),  is given by
\begin{equation}
\Delta x = a(t_{1})f_{k} (\chi) \Delta \theta .
\end{equation}
Thus, the angular diameter  distance is
\begin{equation} \label{add1}
d_{A}  = a(t_{1})f_{k} (\chi) = \frac{a_{0} f_{k} (\chi)}{1+z} 
\end{equation}
and comparing it with Eq. (\ref{dl}) one has
\begin{equation} \label{dual}
d_{A}  = \frac{d_L}{(1+z)^2}, 
\end{equation}
which is called duality relationship.  Eq. (\ref{dual}) is valid beyond the FRW metric. In fact, it is valid 
for any metric in which the flux is conserved. 


\section{The physical Universe \label{phys-uni}}
In the following we provide with theoretical tools to understand the physics of the early Universe.  We treat  
some micro-physics that rules the interactions of the particles and fields.  

\subsection{Thermodynamics in the early Universe }
In the early Universe one considers  a plasma of particles and 
their antiparticles, as originally was done by 
Gamow \cite{Ga46}, who has first considered a physical scenario for the  hot big bang model for
the Universe's beginning. Later on, with the development of modern particle 
physics theories in the 70's it was unavoidable to think about a physical 
scenario which should include  the ``new'' physics for the early Universe.  
It was also realized that the physics described by GR should not be applied 
beyond Planckian  initial conditions, because there the quantum corrections to 
the metric tensor become very important, a theory which is still in progress.  
So the things, one assumes at some early time, $t \, {}^{>}_{\sim} \,  t_{Pl}$,  that 
the Universe was filled with a plasma of relativistic particles  
which include quarks, leptons, and gauge and Higgs bosons, all in thermal 
equilibrium at a very high temperature, $T$, with some gauge symmetry 
dictated by a particle physics theory.  

Theoretically, in order to work in that direction one introduces some thermodynamic 
consi\-de\-rations necessary for the description of the physical content of 
the Universe, which we would like to present here.   Assuming an ideal-gas 
approximation, the number density $n_{i}$ of the particles of type $i$, with a 
momentum $q$, is given by a Fermi or Bose distribution \cite{KoTu90}:

\begin{equation}
n_{i} = \frac{g_{i}}{2 \pi^{2}} 
\int \frac{q^{2} dq}{e^{(E_{i}- \mu_{i})/T} \pm 1} \,\ , 
\label{dissc} \end{equation}
where $E_{i}=\sqrt{m_{i}^{2}+ q^{2}}$ is the particle energy, $\mu_{i}$ is 
the chemical potential, the sign $(+)$ applies for fermions and $(-)$ for 
bosons, and  $g_{i}$ is the number of spin states. One has that 
$g_{i}=2$ for photons, quarks, baryons, electron, muon, tau, and their 
antiparticles, but  $g_{i}=1$ for neutrinos because they are only left-handed.  
For the particles existing in the early Universe one usually assumes 
that $\mu_{i} =0$: one expects that in any particle reaction the $\mu_{i}$ 
are conserved, just as the charge, energy, spin, and lepton and baryon number, 
as well.  For photons, which  can be created and/or annihilated after some 
particle's collisions, its number density, $n_{\gamma}$, must not be conserved 
and its distribution with $\mu_{\gamma}=0, \,\ E=q=\hbar\omega$, reduces to the
Planckian one.  For other constituents, in order to determine 
the $\mu_{i}$, one needs 
$n_{i}$; one notes from Eq. (\ref{dissc}) that for large $\mu_{i}>0, \,\  n_{i}$ is large 
too.   One does not know $n_{i}$, but the WMAP data constrains the baryon density  
at nucleosynthesis such that 
\cite{CyFiOlSk04}:
\begin{equation}
\eta \equiv \frac{n_{B}}{n_{\gamma}} \equiv 
\frac{n_{\rm baryons} - n_{\rm anti-baryons}}{n_{\gamma}} =  
6.14 \pm 0.25 \times 10^{-10} \,\ .
\label{barasc} \end{equation}
The smallness of the baryon number density, $n_{B}$, relative to the 
photon's, suggests that $n_{\rm leptons}$ may also be small compared with  
$n_{\gamma }$.  Therefore, one takes for granted that  
$\mu_{i}=0$ for all particles.  Why the ratio 
$n_{B}/n_{\gamma}$ is so small, but not zero, is one of the puzzles of the 
standard model of cosmology called {\it baryogenesis}, as we will explain below.  

The above approximation allows one to treat the density and pressure of all 
particles as a function of the temperature only. According to the second law 
of thermodynamics, one has \cite{We72}:
$
d S(V,T) = \frac{1}{T}[d(\rho V) + p dV] 
$, 
where $S$ is the entropy in a volume $V\sim a^{3}(t)$ with  
$\rho = \rho(T), \,\ p = p(T)$ in equilibrium. 
Furthermore, the following integrability condition 
$\frac{\partial^{2} S}{\partial T \partial V} = 
 \frac{\partial^{2} S}{\partial V \partial T}$  is also valid, which turns 
out to be
\begin{equation}
\frac{d p}{dT} = \frac{ \rho  + p }{T}   \,\ .
\label{psc}
\end{equation} 
On the other hand, the energy conservation law, Eq. (\ref{frw3asc}), leads to
\begin{equation}
a^{3}(t) \frac{d p}{dt} = \frac{d}{dt}[a^{3}(t)(\rho  + p) ]
\label{cosc}
\end{equation} 
and using Eq. (\ref{psc}), the latter takes the form
$
\frac{d}{dt}[\frac{a^{3}(t)}{T}(\rho  + p) ] =0$,  
and using Eq. (\ref{psc}) again, the entropy equation can be written as
$
d S(V,T) = \frac{1}{T}d[(\rho+p) V] - \frac{V}{T^{2}}(\rho+p) dT 
$.  Last two equations  imply that the entropy is a constant of motion: 
\begin{equation}
S = \frac{a^{3}}{T}[\rho  + p ] = {\rm const.}  \, .
\label{entconsc}
\end{equation}

The density and pressure are given by
\begin{equation}
\rho \equiv  \int E_{i} n_{i} dq \,\ , \qquad   
p  \equiv   \int \frac{q^{2}}{3 E_{i}}  n_{i} dq \,\ .
\label{rpsc}\end{equation}
For photons or ultra-relativistic fluids, $E=q$, these equations become such that  
$p=\frac{1}{3} \rho $,  and then 
confirming Eq. (\ref{ecsc}) with $w = 1/3$, and after integrating 
Eq. (\ref{psc}),  it comes out that
\begin{equation}
\rho = b T^{4} \, , 
\label{bbsc}\end{equation} 
with the constant of integration, $b$.  In a real scenario there are many
relativistic particles present, each of which contributes like Eq. 
(\ref{bbsc}).  
By including all of them, $\rho = \sum_{i} \rho_{i}$ and 
$p = \sum_{i} p_{i}$ over all relativistic species, one has that 
$b(T)=\frac{\pi^{2}}{30} (N_{B} + \frac{7}{8}N_{F} )$, 
which depends on the effective relativistic degrees of freedom of 
bosons ($N_{B}$) and fermions ($N_{F}$); therefore, this quantity varies with 
the temperature; different $i-$species remain relativistic until some 
characteristic 
temperature $T\approx m_{i}$, after that the value $N_{F_{i}}$
(or $N_{B_{i}}$) contributes no more to $b(T)$.  The factor 7/8 accounts for 
the different statistics the particles have, see 
Eq. (\ref{dissc}).  In the standard model of particles physics 
$b\approx 1$ for $T \ll 1$ MeV and $b\approx 35$ for $T>300$ GeV \cite{KoTu90}.  Also for 
relativistic particles, one obtains from Eq. (\ref{dissc}) that
\begin{equation}
n = c T^{3}, \,\ \,\ {\rm with} \,\ 
 c=\frac{\zeta(3)}{\pi^{2}} (N_{B} + \frac{3}{4} N_{F}) \,\ .
\label{nsc} \end{equation}
where $\zeta(3)\approx 1.2$ is the Riemann zeta function of 3. Nowadays, 
$n_{\gamma} \approx \frac{422}{{\rm cm}^{3}} T^{3}_{2.75}$, where
$T_{2.75}\equiv \frac{T_{\gamma_{0}}}{2.75 {}^{\circ}\!K}$.

From Eq. (\ref{entconsc}), and using the relativistic equation of state given above ($w=1/3$), one 
gets that $T\sim 1/a(t)$ and from its solution in Eq. 
(\ref{frwsolsc}) one has, 
\begin{equation}
T = \sqrt[4]{\frac{M_{\frac{1}{3}}}{b}} \frac{1}{a(t)} 
  = \sqrt[4]{\frac{3}{32 \pi G b}} \frac{1}{(t-t_{*})^{\frac{1}{2}}} \,\ ,
\label{tradsc}
\end{equation}
a decreasing temperature behavior as the Universe expands.  Then, 
initially at the big bang $t=t_{*}$ implies $T_{*}= \infty$, the 
Universe was not only very dense but also very hot. 

The entropy for an effective relativistic fluid is given by Eq. (\ref{entconsc})
together with its equation of state and Eq. (\ref{bbsc}), 
$S=\frac{4}{3} ~ b ~ (a ~T)^{3}= {\rm const.}  $
Combining this with Eq. (\ref{tradsc}), one can compute the value of 
$M_{\frac{1}{3}}$ to be 
$M_{\frac{1}{3}}=(\frac{3}{4}S)^{4/3} /b^{1/3}$ 
$\approx 10^{116}$, since $b\approx 35$ and the photon entropy
$S_{0} = \frac{4}{3} ~ b ~ (a_{0} ~T_{0})^{3} \approx 10^{88}$ for the
nowadays evaluated quantities $a_{0} \rightarrow d_{H}(t_{0})=10^{28}$cm and 
$T_{\gamma_{0}}=2.7 ~{}^{\circ}\!K$.  One defines the 
entropy per unit volume, {\it entropy density}, to be 
$s \equiv S/V = \frac{4}{3} \frac{\pi^{2}}{30} (N_{B}+\frac{7}{8}N_{F}) T^{3}$,
then, nowadays $s \approx 7 n_{\gamma}$.  The nucleosynthesis bound on $\eta$,
Eq. (\ref{barasc}), implies that $n_{B}/s \approx  10^{-11}$.

Now  we consider particles in their non-relativistic limit ($m \gg T$).  From
Eq. (\ref{dissc}) one obtains for both bosons and fermions that
\begin{equation}
n= g  \left(\frac{m T}{2 \pi}\right)^{3/2} e^{-m/T} \,\ .
\label{nnrsc} \end{equation}
The abundance of equilibrium massive particles decreases exponentially once
they become non-relativistic; this situation is referred as {\it in equilibrium 
annihilation}.  Their density and pressure are given through 
Eqs. (\ref{rpsc}) and (\ref{nnrsc}) by 
$\rho =n m$ and $ p  = n T \ll \rho$.  
Therefore,  the entropy given by Eq. (\ref{entconsc}) for 
non-relativistic particles, using last two equations,   
diminishes also exponentially during their in equilibrium annihilation.  
The entropy of these particles is transferred to that of 
relativistic components by augmenting their temperature.  Hence, 
the constant total entropy is essentially the same as the one given above, but the $i-$species contributing to it are just those 
which are in equilibrium and maintain their relativistic behaviour, that is,
particles without mass such as  photons.

Having introduced the abundances of the different particle types, we 
would like to comment on the equilibrium conditions for the 
constituents of the Universe, as it evolves.  This is especially of importance 
in order to have an idea whether or not a given $i-$species disappears or 
decouples from the primordial brew.   To see this, let us consider 
$n_{i}$ when the Universe 
temperature, $T$, is such that (a) $T \gg m_{i}$, during the ultra-relativistic
stage of some particles of type $i$ and (b) $T \ll m_{i}$, when the particles 
$i$ are nonrelativistic, both cases first in thermal equilibrium.  From Eq.  
(\ref{nsc}) one has for the former case that $n_{i} \sim T^{3}$; the total 
number of particles, $\sim n_{i} a^{3}$, remains constant. 
Whereas for the latter case, from Eq. (\ref{nnrsc}), 
$n_{i} \sim T^{3/2} e^{-m_{i}/T}$, i.e., 
when the Universe temperature goes down below $m_{i}$, the number density of 
the $i-$species significantly diminishes; it occurs an in equilibrium  
annihilation.  Let us take as example the neutron--proton 
annihilation, one has 
\begin{equation}
\frac{n_{n}}{n_{p}} \sim e^{\frac{m_{p}-m_{n}}{T}} 
= e^{- \frac{1.5 \times 10^{10} \,\  {}^{\circ}\!K}{T}}  \, ,
\label{npsc}\end{equation}
which drops with the temperature, from near to 1 at 
$T \ge 10^{12} \,\ {}^{\circ}\!K$  to about 5/6 at 
$T \approx 10^{11} \,\ {}^{\circ}\!K$, and 3/5 at
$T \approx 3 \times 10^{10} \,\ {}^{\circ}\!K$ \cite{Na02}. If this is forever 
valid, one ends without massive particles, and our Universe should have 
consisted only of radiative components; our own existence prevents that!  
Therefore, eventually the in equilibrium annihilation had to be stopped.  The 
quest is now to freeze out this ratio to be $n_{n}/n_{p} \approx 1/6$ (due to 
neutron decays, until the time when nucleosynthesis begins, $n_{n}/n_{p}$ 
reduces to 1/7) in order to leave the correct number of hadrons for later achieving 
successful nucleosynthesis.  The answer comes from comparing the 
Universe expansion rate, $H$,  with particle physics 
reaction rates, $\Gamma$.  Hence, for $H<\Gamma$, the particles interact 
with each other faster than the Universe expansion rate, then equilibrium is
established.  For $H>\Gamma$ the particles cease
to interact effectively, then thermal equilibrium drops out. This is only approximately 
true; a proper account of that involves a  Boltzmann equation analysis.  
For that analysis numerical integration should be carried out 
in which annihilation rates are balanced 
with inverse processes, see for example \cite{St79,KoTu90}.  In this way, 
the more interacting the particles are, the longer they remain 
in equilibrium annihilation and, therefore, the lower their number 
densities are after some time, e.g., baryons vanish first, then charged 
leptons, neutral leptons, etc.; 
finally, the massless photons and neutrinos, whose particle numbers 
remain constant, as it was mentioned above.  
Note that if interactions of an $i-$species freeze out when it is still 
relativistic, then its abundance can be significant nowadays.

It is worth to mention that if the Universe would expand faster, then the 
temperature of decoupling, when $H \sim \Gamma$, would be higher, then the fixed  
ratio $n_{n}/n_{p}$ must be greater,  and the ${}^{4}$He abundance would be higher, thus 
leading to profound implications in the nucleosynthesis 
of the light elements.  Thus, the expansion rate cannot 
arbitrarily be modified during the equilibrium era of some particles.  Furthermore, if a particle 
species is still highly relativistic ($T \gg m_{i}$) or highly non-relativistic 
($T \ll m_{i}$), when decoupling from primordial plasma occurs, it maintains 
an equilibrium distribution; the former characterized by $T_{r} a = $const. and 
the latter by $T_{m} a^{2} = $const., cf. Eq. (\ref{tmsc}).

There are also some other examples of decoupling, such as  neutrino decoupling:
during nucleosynthesis there exist reactions, e.g. 
$ \nu \bar{\nu} \longleftrightarrow e^{+} e^{-}$, which maintain neutrinos 
efficiently coupled to the original plasma ($\Gamma>H$) until about 1 MeV, 
since $\frac{\Gamma}{H} \approx \left(\frac{T}{{\rm MeV}}\right)^{3}$.  Below
1 MeV reactions are no more efficient and neutrinos decouple and continue 
evolving with a temperature $T_{\nu}\sim 1/a$.  Then, at 
$T \, {}^{>}_{\sim} \, m_{e}=0.51$MeV the particles in equilibrium are photons 
(with $N_{B}=2$) and electron and positron pairs (with $N_{F}=4$) to contribute 
to the entropy with 
$b(T)=\frac{\pi^{2}}{30} (11/2)$.  Later, when the temperature drops to   
$T \ll m_{e}$, the reactions are no more efficient ($\Gamma < H$) and  
after the $e^{\pm}$ pair annihilation there are 
only photons in equilibrium with $b(T)=\frac{\pi^{2}}{30} (2)$.  Since the 
total entropy, $S = \frac{4}{3} b (aT)^{3}$, must be conserved, the decrease in 
$b(T)$ must be balanced with an increase in the radiation temperature, then one
has that $\frac{T_{\gamma}}{T_{\nu}} = \left( \frac{11}{4} \right)^{\! 1/3}$,
which should remain so until today, implying the existence of a cosmic
background of neutrinos with a temperature today of 
$T_{\nu_{0}} = 1.96 \,\ {}^{\circ}\!K$.   This cosmic relic has not been measured yet.

Another example of that is the gravitation decoupling, which should be also 
present if gravitons were in thermal equilibrium at the Planck time and then 
decouple.  The today background of temperature should be characterized at most 
by $T_{\rm grav.} = \left( \frac{4}{107} \right)^{\! 1/3} \approx 0.91 \,\ 
{}^{\circ}\!K$.

For the matter dominated era we have stressed that effectively $p=0$; next we 
will see the reason of this.  First consider an ideal gas (such as  atomic 
Hydrogen) with mass $m$, then $\rho = n m + \frac{3}{2} n  T_{m}$ and 
$p=n  T_{m}$.  From  Eq. (\ref{cosc}) one obtains, equivalently, that 

\begin{equation}
\frac{d}{d a} (\rho a^{3}(t)) = -3 p a^{2}(t)
\label{co1sc}  \end{equation}
and substituting the above $\rho $ and $p$, one has that
\begin{equation}
\frac{d}{d a} (n m a^{3}(t) + \frac{3}{2} n T_{m} a^{3}(t) ) = 
-3 n T_{m} a^{2}(t)
\label{co2sc} \end{equation}
where $n m a^{3}(t)$ is a const. This Eq. yields that 
\begin{equation}
T_{m} a^{2}(t)= {\rm const.} \,\ ,
\label{tmsc} \end{equation} 
the matter temperature drops faster than that of
radiation as the Universe expands, cf. Eq. (\ref{tradsc}).  Now, if one 
considers both radiation and matter, one has that 
$\rho = n m + \frac{3}{2} n  T_{m} + b T_{r}^{4}$ and 
$p=n  T_{m} + \frac{1}{3} b T_{r}^{4}$; the source of Universe's expansion 
is proportional to 
$\rho +3 p =  n m + \frac{9}{2} n  T_{m} + 2  b T_{r}^{4}$, 
the first term dominates the second, precisely because $T_{m}$ decreases 
very rapidly. The third term diminishes as $\sim 1/a^{4}$, whereas the first as 
$\sim 1/a^{3}$, and after the time of densities equality, 
$\rho_{m} = \rho_{r}$, 
the matter density term is greater than the others, that is why one assumes no 
pressure for that era.  

From now on, when we refer to the temperature, $T$, it should be related to 
the radiation temperature.  The detailed description of the Universe thermal evolution for the different 
particle types, depending on their masses, cross-sections, etc.,  
is well described in many textbooks, going from the physics 
known in the early 70's \cite{We72} to 
the late 80's \cite{KoTu90}, and therefore it will not be presented here.  
However, we notice that as the Universe cools down a series of
spontaneous symmetry--breaking (SSB) phase transitions are expected to occur. 
The type and/or nature of these transitions depend on the specific particle
physics theory considered.  Among the most popular ones are Grand
Unification Theories (GUT's), which bring together all known interactions 
except for gravity.  One could also be more modest and just consider the 
standard model of particle physics or some extensions of it.  Ultimately, 
one should settle, in constructing a cosmological theory, up to which energy
scale one wants to describe physics. For instance, at a temperature between 
$10^{14}$ GeV to $10^{16}$ GeV the transition of the $SU(5)$ GUT should took  
place, if this theory would be valid, in which a Higgs field breaks this symmetry
to $SU(3)_{C} \times SU(2)_{W} \times U(1)_{HC}$, a process through which
some bosons acquired their masses.  Due to the gauge symmetry, there
are color (C), weak (W) and hypercharge (HC) conservation, as the subindices  
indicate.  Later on, when the Universe evolved to around 150 GeV the 
electroweak phase transition took place in which the standard model  Higgs field broke 
the symmetry $SU(3)_{C} \times SU(2)_{W} \times U(1)_{HC}$ to   
$SU(3)_{C} \times U(1)_{EM}$; through this breaking fermions 
acquired their masses.  At this stage, there were only color and electromagnetic
(EM) charge conservation, due to the gauge symmetry.  
Afterwards, around a temperature of 175 MeV the Universe should underwent 
a transition associated to the chiral symmetry--breaking  and color confinement from which baryons
and mesons were formed out of quarks.   Subsequently,  at approximately 
10 MeV begun the synthesis of light elements (nucleosynthesis), when most
of the today observed Hydrogen, Helium, and some other light elements 
abundances were produced.  So far the nucleosynthesis represents the earliest scenario
tested in the standard model of cosmology.  After some thousand years ($z \sim 3200$ \cite{wmap7}), the Universe 
is matter dominated, over the radiation components.  At about 380, 000 years ($z \sim 1090$ \cite{wmap7}) 
recombination took place, that is,  the Hydrogen ions and electrons combined to compose neutral Hydrogen atoms, 
then matter and EM radiation decoupled from each other; at this moment (baryonic) matter structure begun to form.  Since 
that moment  the surface of last scattering of the CMBR evolved as an imprint of the early 
Universe. This is the light that Penzias and Wilson first measured, and was later measured in more 
detail by BOOMERANG, MAXIMA, COBE, and WMAP, among other probes.  PLANCK cosmological 
data will be forthcoming in 2012. 

\subsection{Inflation: the general idea \label{giin}} 

As we mentioned above, the FRW cosmological Eqs.
(\ref{frw1sc})-(\ref{frw3asc}) admit very rapid expanding solutions for the
scale factor.  This is achieved when  $\rho +3p$,
is negative, i.e., when the equation of state admits negative pressure such
that $w <-1/3$, to have $\ddot{a}>0$.  For instance, if $w =-2/3$, one has
that $a \sim t^{2}$ and $\rho \sim 1/a$, that is, the source of rapid expansion
decreases inversely proportional with the expansion.  Of special interest is
the case when $w =-1, ~ \rho ={\rm const.}$, because this
guarantees that the expansion rate will not diminish.  Thus, if
$\rho ={\rm const.}$ is valid for a period of time, $\tau$,  the Universe
will experience an expansion of $N=\tau H$ foldings, given by
$a=a_{*} e^{N}$, Eq. (\ref{isc}).  This is the well known de Sitter 
cosmological solution \cite{deSi17},  achieved here only for a $\tau$-stage in a FRW model.

We shall now see how an inflationary stage helps to solve the horizon and flatness 
problems of the old standard cosmology.  Firstly consider the particle (causal) horizon, given by
Eq. (\ref{h1sc}), during inflation, again with $k=0$, one obtains
\begin{equation}
d_{H} = H^{-1} (e^{Ht} -1) \,\ ,
\label{hin} \end{equation}
the causal horizon grows exponentially, whereas $H^{-1}$ remains constant.  We 
compare the horizon distance with that of any physical length scale, 
$L(t)= L_{*} \frac{a(t)}{a_{*}}= L_{*} e^{Ht}$, to get
\begin{equation}
\frac{d_{H}}{L} = \frac{H^{-1}(e^{Ht} -1)}{L_{*} e^{Ht}} \, {}^{>}_{\sim} \,
1 - e^{-Ht}   \,\ ,
\label{hlin} \end{equation}
for initial length scales $L_{*} \, {}^{<}_{\sim} \, H^{-1}$.  After a few
e-fold times the causal horizon is as big as any length scale that was  initially 
subhorizon sized.  Therefore, if the original patch before inflation is
causally connected, and presumably in equilibrium, then after inflation this region of causality is
exponentially bigger than it was, and all the present observed (apparent)
Universe can stem from it,  solving the horizon
problem.  In fact, if the inflation stage is sufficiently large, there can
exist nowadays regions which are so distant away from each other that
they are still not in contact, even though originally they come from the same
causal patch existing before inflation. These regions will be for a time not in contact 
since we currently are experiencing an accelerated expansion, and then, if this is of exponential 
type, the event horizon is constant  and light/information that shall come to us will be 
from only a delimited region $H^{-1}$, as we explain below.

From Eq. (\ref{hlin}) one can observe that if the initial physical length
scale is greater than the Hubble distance, $L_{*}>H^{-1}$, then $d_{H}< L$
during inflation.  Events initially outside the Hubble horizon remain acausal.
This is better seen by considering the event horizon, $d_{e}$,  defined in 
Eq. (\ref{eh1in}).  This delimits the region of space which will keep in causal contact after some
time;  that is, it delimits the region from which one can ever receive (up to
some time $t_{\rm max}$) information about events taking place now (at the time $t$).  
During inflation one has that
\begin{equation}
d_{e} = H^{-1} (1 - e^{-(t_{\rm max}-t) H}) \approx H^{-1},
\label{eh2in} \end{equation}
which implies that any observer sees only those events that take place within
a distance $\le H^{-1}$.  In this respect, there is an analogy with black
holes, from whose surface no information can get away.  Here, in an exponential
expanding Universe, observers encounter themselves in a region
which apparently were surrounded by a black hole \cite{GiHa77,Li90}, since they
receive no information located farther than $H^{-1}$.

The apparent horizon at present stems from a region
delimited by the original patch $d_{e}  \approx H^{-1}$, which during inflation remains almost
constant and, afterwards, evolves as $H^{-1}\sim t$.  At the end of inflation
$a(t)\gg H^{-1}(t)/H_{0}^{-1}$.  Subsequently, the scale factor expands only with the
power law solution $t^{1/2}$ (and later as $t^{2/3}$), whereas the Hubble horizon
evolves faster, $H^{-1}\sim t$.  Then, at some later time the Hubble horizon is
as large as the scale factor, $H^{-1}\sim a(t) H_{0}^{-1}$.  Accordingly, there is  a 
minimal number of efolds of inflation, $N \sim 60$,  necessarily to have this equality at present (this number depends on the 
energy scale of inflation, see for instance J. L. Cervantes-Cota in  \cite{BrCeSa04}); that is, 
the original patch grown until now is as big as our apparent, Hubble horizon.  Hence, some time ago,
say, at the last scattering surface (photon decoupling) the Universe consisted
of $10^{5}$ Hubble horizon regions, yet all these regions stem from one
original patch of size $H^{-1}_{*}$ at the start of inflation.  

A  typical scale $L_{*} \le H^{-1}$ will increase exponentially its size as
$L(t)= L_{*} \frac{a(t)}{a_{*}}= L_{*} e^{N}$.  That is, all physical
inhomogeneities, anisotropies and/or `perturbations'  of any kind (including
particles!) will be diluted away from a region $d_{e} \sim H^{-1}$, and its density becomes insignificant, thus
solving the monopole (and other relics) problem.

On the other hand, the flatness problem in the old standard cosmology arises since  $\Omega$
approaches closely to unity as one goes back in time in a way that one has to 
choose very special initial density values, at the Planck time $\Omega_{Pl} -1 \approx \pm 10^{-59}$, for 
explaining our flatness today, i.e.,
$\Omega_{0}\approx {\cal O}(1)$.  Now,  imagine the Universe with 
initial conditions such that $\Omega_{*} -1 \approx k$.
Now, if the exponential expansion  occurs, $\Omega(t=\tau)$ evolves to
\begin{equation}
\Omega(\tau) -1 =\frac{\rho-\rho_{c}}{\rho_{c}}  =
\frac{k}{a^{2} H^{2}} =  k e^{-2N} \,\ .
\label{oin} \end{equation}
If $N$ is sufficiently large, which will be case since typically
$N > 60$, the Universe looks after a de Sitter stage like an almost
perfect flat model.  Therefore, it plays almost no role what the initial
density was, if the exponential expansion occurs -guaranteed by the cosmological no hair- 
the Universe becomes effectively flat.   In this way, 
instead of appealing to very special initial conditions,  one starts with
an Universe with more normal conditions, that is, non fine tuned, which
permit the Universe to evolve to an inflationary stage, after which it looks
like it would had very special conditions, i.e.,  with $\Omega\approx 1$ with
exponential accuracy.  

After inflation the Universe contains a very small particle density and is very cold, even as cold
as the CMBR is today!  The transition to a radiation dominated era with
sufficient entropy and particle content comes from the `decaying' or
transformation of the energy source of inflation, $\rho = V(0)$, into heat; a
process called {\it reheating} (RH).

\subsection{Reheating and baryogenesis}

At the end of inflation the $\phi$-field (the inflaton) begins to oscillate around its stable, global
minimum, say $v$.   Its oscillation frequency is given by the effective mass of the Klein Gordon equation 
in a FRW Universe, $\ddot{\phi} + 3 H \dot{\phi} + V'(\phi) = 0$. 
After inflation the term $V'(\approx V'' \phi=M^{2}~ \phi)$ is greater than
$3H\dot{\phi}$, since  the Hubble rate evolves hereafter always decreasing,
$H\sim 1/t$.  Thus, the oscillation frequency of the $\phi$-field is simply given
by the field mass, $M$.   The stored energy of the inflaton 
field, $\rho_{H}=V+\frac{1}{2} \dot{\phi}^{2}$, can decay to give rise to
quantum  particle creation \cite{Rh82}.  The reheating models depend on the particle 
physics models but general features of the process have been understood.  We firstly explain the old scenario 
called reheating and secondly the {\it preheating} that seems to be more realistic.   In reheating, the 
state $\phi=v$ is considered as a coherent state of scalar particles in 
rest.  Then, this state decays through the ordinary decay of the field bosons 
and the decay rate coincides with the rate of decrease of the energy of 
oscillations.  Thus, the decay rate of the boson, $\Gamma_{H}$, 
introduces a friction term of the type $\Gamma_{H} \dot{\phi}$ in the Klein Gordon equation 
that causes the scalar field to vanish and the reheating of the
Universe \cite{Rh82}.  The transformed energy goes into masses and kinetic
energy of the new particles produced.  Typically, the produced particles
(bosons, fermions) have smaller masses than the field boson, therefore much of
their energy goes into kinetic energy, and particles behave as a relativistic
fluid.  If the decay rate is greater than the Hubble rate after inflation,
$\Gamma_{H}>H(t=t_{f})\equiv H_{f}$, then the reheating process occurs within
an expansion time, very rapidly.  In this  case, the reheating temperature is
\cite{StTu84} $T_{RH} \approx \left( \frac{\rho_{H}}{b} \right)_{f}^{1/4} =
\left(\frac{V + \frac{1}{2} \dot{\phi}^{2}}{b} \right)_{f}^{1/4}$, where 
$b=\frac{\pi^{2}}{30} (N_{B} + \frac{7}{8}N_{F} )$, $N_{B} ~(N_{F})$
stands for boson (fermion) degrees of freedom.  For high temperatures,  $T>300$
GeV, one has that $b\approx 35$, see discussion after Eq. (\ref{bbsc}).  The
subindex $f$ means to be evaluated  at the end of inflation.  There, the
kinetic and potential energies are of the same order of magnitude.  Then, for a 
self-interaction  potential  with 
$V\approx V(0) = \frac{1}{24} \lambda v^{4}$, one has that
\begin{equation}
T_{RH} \approx \left( \frac{\lambda v^{4}}{24 b}\right)^{1/4} =
\frac{\sqrt{M \,  v}}{\sqrt[4]{8 b}} \,\ .
\label{tr2in} \end{equation}

One the other hand, if the decay process is rather slow,  $\Gamma_{H}<H_{f}$,
then the  field continues to oscillate coherently until
$t=\Gamma_{H}^{-1}$.  During this time the solution of the Klein Gordon equation is
$\phi\sim \frac{1}{t} \cdot {\rm cos}M \, t$, $H=\frac{2}{3t}$.  The coherent
oscillations behave  as non-relativistic matter fluid ($w=0$), i.e., 
$a\sim t^{2/3}~$ \cite{Tu83}.  If field bosons do not 
completely decay, the oscillations represent a ``sea'' of cold bosons with 
$M \gg T$. They can account for the cold dark matter, but some 
degree of fine tuning is necessary \cite{LiUr06}. 

Without partial or total decaying of field oscillations the Universe
remains cold and devoid  of fermions and (other) bosons.  Therefore, let us
suppose that indeed reheating took place, but now with
$\Gamma_{H} \, {}^{<}_{\sim} \, H_{f}$, then the reheated temperature is
$T_{RH} \approx (\sqrt{M v}/(\sqrt[4]{8 b}) \, 
\sqrt{\Gamma_{H}/H_{f}} $,  
a factor $\sqrt{\Gamma_{H}/H_{f}}$ smaller than the efficient reheating
case, Eq. (\ref{tr2in}).   The reheating process occurs normally within one or few Hubble times.  Then,
the scale factor does not increase significantly during it. 

The reheating scenario presented above is based on the original theory
developed in the context of the new inflationary scenario, however, it is also
applicable to other models.  In the course of the time important steps to
consolidate the theory were made, see for example Ref. \cite{TrBr90}. But 
qualitative new ideas were introduced in Refs. \cite{KoLiSt94-96}.  Accordingly, the 
process of reheating should consist of three different stages.  At the first phase, the $\phi$-field decays into
massive bosons (fermions) due to a parametric resonance given through a Mathieu
equation that determines the regions of stability and instability (particle
production) of the quantum fluctuations of the created particles.  These can be
$\phi$-particles or other bosons (fermions) coupled to the $\phi$-field.  This
process is very efficient, even explosive, and much bosons can be created in
this stage.  Note that the original theory is based upon the decay of the
$\phi$-particles, whereas in the present theory the $\phi$-field decays  into
$\phi$-particles, and perhaps others, and only after this process the decay of
these particles proceeds.  Then, to distinguish this explosive process from the
normal stage of particle decay, the authors of Ref. \cite{KoLiSt94-96} called
it {\it preheating}.  Bosons produced at this stage are far away from thermal
equilibrium and have very big occupational numbers.  The second stage of this
scenario describes the decay of the already produced particles.  This phase is
described as in the original reheating theory.  Then, the methods
developed for the original theory are now applied to the product particles, but
not itself to the decay of the $\phi$-field.  The third stage is the
thermalization by which the system reaches equilibrium \cite{FeKo01}.

The process of reheating is very complex and depends fine on  the particle
physics theory one has in turn.  As a matter of fact, one expects a reheat
temperature $T_{RH}>$ few MeV to be able to attain nucleosynthesis.  A second,
and more restrictive constraint comes from  {\it baryogenesis}.  One can see this by
noting that the number density of any conserved quantity before reheating
divided by the entropy density $\frac{n}{s}$ becomes after reheating
insignificant because of the huge entropy produced.  One gets
$\frac{n}{s} |_{{}_{rf}} =e^{-3N}\frac{n}{s} |_{{}_{ri}}$, where $ri$ and $rf$ denote
the ini\-tial and final state of reheating, respectively.  In this way, any
baryon asymmetry initially present will be brought to unmeasurable values.  Therefore,
after reheating the baryon asymmetry must be created.  Note also that any
unwanted relic (${\rm x}$), accounted through $\frac{n_{{\rm x}}}{s}$, will
essentially disappear after reheating.  The correct  baryon-antibaryon balance at nucleosynthesis is 
$ n_{B}/s \approx  10^{-11}$,  in consistency with Eq. (\ref{barasc}).  There are  some attempts 
to achieve baryogenesis  at low energy scales, as low as few GeV or TeV \cite{low-scale-baryo90s}.  
Recent attempts to solve this problem seek to yield a prior a lepton asymmetry, {\it leptogenesis},  generated 
in the decays of a heavy sterile neutrino \cite{DaNaNi08},  to later end with baryogenesis. 

\section{The Perturbed Universe}
In the previous sections we have outlined how the evolution of a homogeneous Universe can be described by 
means of few equations and simple concepts such as the ideal perfect fluids.
The next step is introducing in this scenario small inhomogeneities 
that can be treated as first order perturbations to those equations, 
the goal being the description of the structures we see today in the Universe. 
This perturbative approach is sufficient to accurately describe the  small temperature anisotropies ($\Delta T/T\sim10^{-5}$) observed 
in the CMBR today, 
but can describe the distribution of matter today only at those scales that are still in the linear regime. 
At the present epoch, scales smaller than $\sim 30\;{\rm Mpc}\; h^{-1}$ \cite{reid10} have 
already entered the non linear-regime ($\Delta\rho/\rho>>1$) 
due to the fact that matter tends to cluster under the effect of gravity. 
These scales can therefore only be described by means of numerical or semi-numerical approaches \cite{carlson09}.

The approach is quite straightforward but involves a differential equation for the density perturbation 
of each individual constituent: scalar fields in inflation, or baryons, radiation, neutrinos, DM, and 
DE (usually treated as cosmological constant) in later times, and in general needs to be solved numerically.
In the context of geometry metric and/or GR the metric is treated as the general expansion term 
$g_{\mu\nu}^{(0)}$ plus perturbation $h_{\mu\nu}$:
\begin{equation}
\label{hmn}
g_{\mu\nu}=g_{\mu\nu}^{(0)}+h_{\mu\nu}, 
\end{equation}
with $h_{\mu\nu}<<g_{\mu\nu}^{(0)}$ where $^{(0)}$ indicates the unperturbed homogeneous quantities.

Inhomogeneities in the distribution of the components of the Universe are a source of scalar perturbations 
of the metric.  Nevertheless  vector or tensor perturbations can modify the metric as well.  The standard 
cosmological model do not predict vector perturbations, that would introduce off-diagonal terms in the metric tensor. These 
perturbations would produce vortex motions in the primordial plasma that are expected to rapidly decay. Models with topological 
defects or inhomogeneous primordial magnetic fields instead predict a consistent fraction of vector 
perturbations \cite{seljak97b,turok97,kim09}.

On the other hand, the standard cosmological model predicts the production of gravitational waves during the epoch of inflation, when the Universe 
expanded exponentially, as we will see below.  Gravitational waves induce tensor 
perturbations  $h^T_{\mu\nu}$  on the metric of the type:
\begin{displaymath}
h^T_{\mu\nu} = a^2   
\left( \begin{array}{cccc}
0 & 0 & 0 & 0 \\
0& h_+ & h_{\times} & 0 \\
0& h_{\times} & -h_+ & 0 \\
0 & 0 & 0 & 0
\end{array} \right)
\end{displaymath}
where $h_+$ and $h_{\times}$  are the polarization directions of the gravitational wave. This tensor is traceless, symmetric 
and divergentless, i.e. it perturbs the time space orthogonally to the  propagation direction of the wave.  The amplitude of these 
tensor perturbations are expected to be small compared to scalar ones, and therefore  negligible in first approximation 
as far as we are interested in studying the perturbations of the metric tensor.   Nevertheless these waves are expected to leave 
an imprint in the polarization of the CMBR, and their eventual detection would unveil an extremely rich 
source of information about an epoch of the Universe that is very hardly observable otherwise.

It is important to underline that choosing to model the metric perturbations corresponds to choosing a \emph{gauge}, i.e. a specific 
coordinate system in which the metric tensor is represented. Changing the coordinate system of course do not change the physics, but can 
remarkably vary the difficulty of the calculations and the understanding of the physical meaning of the different quantities. 
To solve the perturbed equations one chooses convenient gauges for the different expansion epochs and 
depending on whether the formalism is theoretical or numerical, as we will see below.

The presence of weak inhomogeneous gravitational fields introduce small perturbations in the metric tensor that can be 
modeled by introducing two scalar functions $\Phi(\vec{x},\eta))$  and $\Psi(\vec{x},\eta)$ \cite{MuFeBr92} in the Robertson-Walker metric as:
\begin{equation}
 ds^2= a^{2} (\eta)\,  \left[ -[1+2\Phi(\vec{x},\eta)] \, d\eta^2+ [1+2 \Psi(\vec{x},\eta)] dx_{i} dx^{i} \right] , 
\end{equation}
where therefore the perturbed part of the metric tensor is: 
\begin{equation}
 h_{00}(\vec{x},\eta) = -2\Phi(\vec{x},\eta), \quad h_{0i}(\vec{x},\eta) = 0, \quad 
h_{ij}(\vec{x},\eta) = a^2\delta_{ij}(2\Psi(\vec{x},\eta)) .
\end{equation}

This metric is just a generalization of the well known metric for a weak gravitational field usually presented in text books (e.g. Chapt. 18 in 
Misner \cite{MiThWh73}) for the case of a static Universe ($a(\eta)=1$).   The function $\Phi$ describes Newton's gravitational field, while 
$\Psi$ is the perturbation of the space curvature. The above gauge  is the \emph{Newtonian conformal gauge}, which has the advantage 
of having a diagonal metric tensor $g_{\mu\nu}$ in which the coordinates are totally fixed with no residual gauge modes and therefore 
with a straightforward interpretation of the functions introduced \cite{MuFeBr92,ma94,ma95}. An example of an alternative gauge 
particularly popular in literature is the \emph{synchronous gauge}, generally defined as:
\begin{equation}
 ds^2=a^2 (\eta) [-d\eta^2+(\delta_{i,j}+h_{i,j}) \, dx^{i} \, dx^{j}] ,
\end{equation}
which is especially used in codes computing the anisotropies and inhomogeneities in the Universe, as better behaved numerically by choosing 
that observers fall freely without changing their spatial coordinates.  A further analysis is found in e.g.  \cite{LyLi09}.  
 
\subsection{Perturbations during inflation \label{denpertin}}
The primeval fluctuations are thought to be present at the very beginning of time, at the inflationary epoch.  The generation 
of perturbations are produced by quantum fluctuations of the $\phi$-field during the accelerated stage, for a review 
see \cite{MuFeBr92,BrCeSa04,LyLi09}.   These fluctuations are usually studied in the {\it comoving gauge} in which the scalar 
field is equal to its perturbed value at any given time during inflation and, therefore, the perturbation information 
resides in the metric components. The perturbations cross outside the event horizon during inflation and re-enter into the horizon
much later, at the radiation and matter dominated epochs, to yield an almost scale invariant density 
perturbation spectrum (Harrison-Zel'dovich, $n_{S} =1$), as the required for structure formation. 

We introduce this topic by noting that the event horizon during a de Sitter
stage is $d_{e}\approx H^{-1}$, cf. Eq. (\ref{eh2in}).  This means that
microphysics can only operate coherently within distances at most as big as
the Hubble horizon, $H^{-1}$.  Recall that the causal horizon, $d_{H}$, expands
exponentially and it is very large compared to the almost constant $H^{-1}$
during inflation, see Eq. (\ref{hin}).  Hence, during the de Sitter stage the
generation of perturbations, which is a causal microphysical process, is
localized in regions of the order of $H^{-1}$.  

It was shown that the amplitude of inhomogeneities produced corresponds to the
Hawking temperature in the de Sitter space, $T_{H}=H/(2\pi)$.  In turn, this means that
perturbations with a fixed physical wavelength of size $H^{-1}$ are produced
throughout the inflationary era.   Accordingly, a physical scale associated to
a quantum fluctuation, $\lambda_{{\rm phys}} = \lambda a(t)$, expands
exponentially and once it leaves the event horizon, it behaves as a metric
perturbation; its description is then classical, general relativistic.  If
inflation lasts for enough time, the physical scale can grow as much as a
galaxy or horizon sized perturbation.  The field fluctuation expands always
with the scale factor and after inflation, it evolves according to $t^{n}$
($n=1/2$ radiation or $n=2/3$ matter).  On the other hand, the Hubble horizon
evolves after inflation as $H^{-1}\sim t$.  This means, it will come a time at
which field fluctuations cross inside the Hubble horizon and re-enters as 
density fluctuations.  Thus, inflation produces a gross spectrum of
perturbations, the largest scale originated at the start of inflation with a
size $H^{-1}_{i}$, and the smallest with $H^{-1}_{f}$ at the end of inflation.
The power spectra for scalar ($S$) and tensor ($T$) perturbations are given by:
\begin{equation}
P_{S} (k) \approx  \left( \frac{H^{2}}{16 \pi^{3}\dot{\phi}_{c}^{2}}\right)
\;\vline{\atop{\atop {}_{k= a H}}},   \qquad 
P_{T} (k) \approx  \left( \frac{H^{2}}{4 \pi^{2} m_{Pl}^{2} }\right)
\;\vline{\atop{\atop {}_{k= a H}}} , 
\label{dcp1in} \end{equation}
where $\dot{\phi}_{c}$ is the classical field velocity. The equations are evaluated at the horizon 
crossing ($k= a H$) during inflation. 
Each of the $k-$modes generate also an anisotropy pattern in the CMBR that was measured for scalar 
perturbations by the COBE \cite{Sm92} and later probes.  The PLANCK satellite may have the chance 
to detect  the ratio of tensor to scalar amplitudes
$r\equiv C_{l}^{T}/ C_{l}^{S} < 0.36$ \cite{wmap7}, since the tensor modes modulate CMBR photons coming from last 
scattering.   Associated to these perturbations one has the spectral indices, $n_S \approx 0.96$ \cite{wmap7} and $n_T$, and 
their runnings, $d n_{S}/d lnk \approx -0.034 \, $ \cite{wmap7}.         

These density and other metric perturbations are small, but we discuss in the next section how to include them
so that the information contained  can be recognized and exploited.

\subsection{Perturbations inside the horizon}
In the early Universe, baryons were tightly coupled to photons in an expanding background. 
Baryonic matter and dark matter potential wells provoked the local collapse of density fluctuations up certain point,
at which the radiation pressure was big enough to pull out the matter apart, and smoothing the potential 
wells.  These oscillations of the plasma  can be thought of as {\it acoustic waves}. As 
we know any wave can be decomposed into a 
sum of modes with different wave numbers, $k = 2\pi/\lambda$.
Since  these modes are in the sky,  their wavelengths  are
measured as angles rather than as distances. 
Accordingly, instead of decomposing the wave in a Fourier series, what is normally done is
to decompose the wave in terms of spherical harmonics, $Y_{lm}(\hat{p}) $. 
The angular power spectrum can be expanded in Legendre polynomials,
since there is no preferred direction in the Universe and that only angular separation $\theta$ is relevant.
 A mode $l$  plays the same role of the wavenumber $k$, thus $l \approx 1/\theta$. Ultimately, we are
interested in the temperature fluctuations that are analyzed
experimentally in pairs of directions $\hat n$ and $\hat n'$ where $cos(\theta) = \hat n \cdot \hat n'$. 
We then average
these fluctuations obtaining the multipole expansions: 
\begin{equation} \label{deltaT}
\frac{\Delta T}{T} = \sum_{l=1}^{\infty}\sum_{m=-l}^{l} a_{lm}(\vec{x},\eta)Y_{lm}(\hat{p}) , 
\qquad 
P_{S}(\theta) = \sum \frac{(2 l +1)}{4 \pi} C_l P_l ({\rm cos}
\theta) , 
\end{equation}
where $P_{S}(\theta)$ is the angular power spectrum, $P_l$ are the Legendre polynomials and $C_l$ is estimated as the average 
over $m$ of $a_{l m}$. All this information can be used to determine the cosmological
parameters $\Omega_i$. We will not discuss
detailed calculations nor the curve that must be adjusted to obtain
the best fit values for such parameters. 
The peak of the fundamental mode appears at approximately
\begin{equation} \label{ele}
l \simeq  \frac{200}{\sqrt{\Omega}} .
\end{equation}

BOOMERANG \cite{deB00} and MAXIMA \cite{Ha00} were 
two balloon-borne experiments designed to measure the
anisotropies at smaller scales than the horizon at decoupling ($\theta_{\rm hor-dec} \sim 1  {}^{\circ}$), hence measuring the 
acoustic features of the CMBR. The sensitivity of the instruments  allowed a
measurement of the temperature fluctuations of the CMBR over a broad
range of angular scales.   BOOMERANG found a value of  $l = 197 \pm 6$ and MAXIMA-1 found a
value of $ l \approx 220$. This implies that the cosmological density parameter $\Omega \approx 1$, see Eq. (\ref{Omega}), implying 
that the Universe is practically flat, $\Omega_{k} \approx 0$.  These two experiments provided with the 
first strong evidence for a flat Universe  from observations. Happily, this result was expected from 
Inflation. These results were confirmed by the 
Wilkinson Microwave Anisotropy Probe (WMAP)  in a series of data releases in the last decade, as well as by other 
cosmological probes:  the Universe is flat or pretty close to be flat.  The problem in the exact determination of the curvature is
because the CMBR anisotropies  show strong degeneracies among the cosmological parameters \cite{BoEfTe97-ZaSpSe97}.
However, the satellite PLANCK will offer results of the density parameters with uncertainties less than a percentage level. 

Since baryons and photons were in thermal equilibrium until recombination (also called {\it last scattering}), the 
acoustic oscillations were  imprinted too in the matter perturbations, as they were in the CMBR  
anisotropies.  These are known as {\it baryon acoustic oscillations} (BAO).  The sound horizon 
at the moment when the baryons decoupled from photons plays a crucial role in determination 
of the position of the baryon acoustic peaks. This time is known as {\it drag epoch} which 
happens at $z_{d} =a_{0}/a_{d} -1$. The sound horizon at that time is defined in terms of the  effective 
speed of sound of the baryon-photon plasma, $c_{s}^{2}  \equiv \delta p_\gamma/ (\delta \rho_{\gamma} + \delta \rho_{b}) $,
\begin{equation}
r_{s} (z_{d}) = \int_{0}^{\eta_{d}} d \eta \, c_{s} (\eta)  = \frac{1}{3} \int_{0}^{a_{d}}   
\frac{da}{a^{2} H(a) \sqrt{1+(3\Omega_{b}/4 \Omega_{\gamma})a}}  \, .
\end{equation}
Note that {\it drag epoch}  does not coincide with last scattering. In most scenarios $z_{ \rm d} < z_{ls}$ \cite{HuSu96}.  
The redshift at the drag epoch can be computed with a fitting formula that is a 
function of $\Omega_{m}^{(0)} h^{2}$ and  $\Omega_{b}^{(0)} h^{2}$  \cite{EiHu98}.  The WMAP -5 year team 
computed these quantities for the $\Lambda$CDM model obtaining
 $z_{d} = 1020.5 \pm 1.6$ and  $r_{s} (z_{d}) = 153,3 \pm 2.0$ Mpc \cite{wmap5}. 

What one measures is the angular position and the redshift \cite{SeEi03,AmTs10}: 
\begin{eqnarray}
\theta_{s} (z) &=& \frac{r_{s} (z_{d}) }{(1+z) \, d_{A}(z)}, \\
\delta z_{s} (z) &=&  r_{s} (z_{d}) \,  H(z), 
\end{eqnarray}
where $d_{A}(z)$ is the proper (not comoving) angular diameter distance, Eq. (\ref{add1}), and $H(z)$ by 
Eq. (\ref{hz}). The angle $\theta_{s} (z) $ corresponds to the direction 
orthogonal to the line-of-sight, whereas $\delta z_{s} (z)$ measures the fluctuations along the line-of-sight.   Observations 
of these quantities are encouraging to determine both $d_{A}(z)$ and $H(z)$.   However, from the 
current BAO data is not simple to independently measure these quantities. This will certainly happen in forthcoming surveys 
\cite{BigBoss11}. Therefore, it is convenient to  combine the two orthogonal 
dimensions to the line-of-sight with the dimension along the line-of-sight to define \cite{Ei05}:
 \begin{equation}
D_{V}(z) \equiv \left( (1+z)^{2} d_{A}(z)^2 \frac{z}{H(z)} \right)^{1/3}  \, ,
\end{equation}
where the quantity  $D_{M} \equiv d_{A}/a = (1+z) d_{A}(z) $ is the comoving angular diameter distance.  One also defines the BAO distance
 \begin{equation}
r_{\rm BAO}(z) \equiv  r_{s} (z_{d}) /  D_{V}(z)  . 
\end{equation}

The BAO signal has been measured in large samples of luminous red galaxies from the SDSS \cite{Ei05}.  There is  a 
clear evidence ($3.4 \sigma$) for the acoustic peak at $100 h^{-1}$ Mpc scale. Moreover, the scale and 
amplitude of this peak are in good  agreement with the prediction of the $\Lambda$CDM given the WMAP data. One finds that 
$D_{V}(z=0.35)= 1370 \pm 64$ Mpc, and more recently new determinations of the BAO signal has been 
published \cite{Ca11} in which $\theta_{s}  (z=0.55)= 3.90{}^{\circ} \pm 0.38{}^{\circ} $, and $w= -1.03 \pm 0.16$ for the equation 
of state parameter of the dark energy, or $\Omega_{M} = 0.26 \pm 0.04$ for the matter density, when the other 
parameters are fixed.  

Measuring the BAO feature in the matter distribution at different redshifts will help to break the degeneracy that exists
in the determination of the cosmological parameters. And by combining line-of-sight with angular determinations of the BAO feature 
one will constrain even more the parameter space. Further, a complete combination of BAO, the full matter power spectrum,  
Supernovae Ia, and CMBR data shall certainly envisage the true nature of the mysterious, dark Universe.     


Finally, we show recent plots of the CMBR power spectrum from the  WMAP 7-year data and from the South Pole Telescope, 
from Ref.  \cite{Ke11}, and  the mass variance ($\Delta_{M}/M = \sqrt{P(k)k^3/(2\pi^2)}$ )  of the reconstructed matter power spectrum from 
the Atacama Cosmology Telescope and other observations, from Ref. \cite{Hl11}.  The left panel shows the extraordinary fit of the 
$\Lambda$CDM model and the importance of foregrounds  for large $l$-modes. The right panel shows the variance decrease as the 
mass increases, covering ten orders of magnitude in the range of masses.  We also notice the effect of BAO at intermediate scales  
and damping on the essentially scale invariant perturbations that one anticipates from inflation. These observations fit remarkably well 
to the $\Lambda$CDM model.

\begin{figure}
\includegraphics[width=80mm]{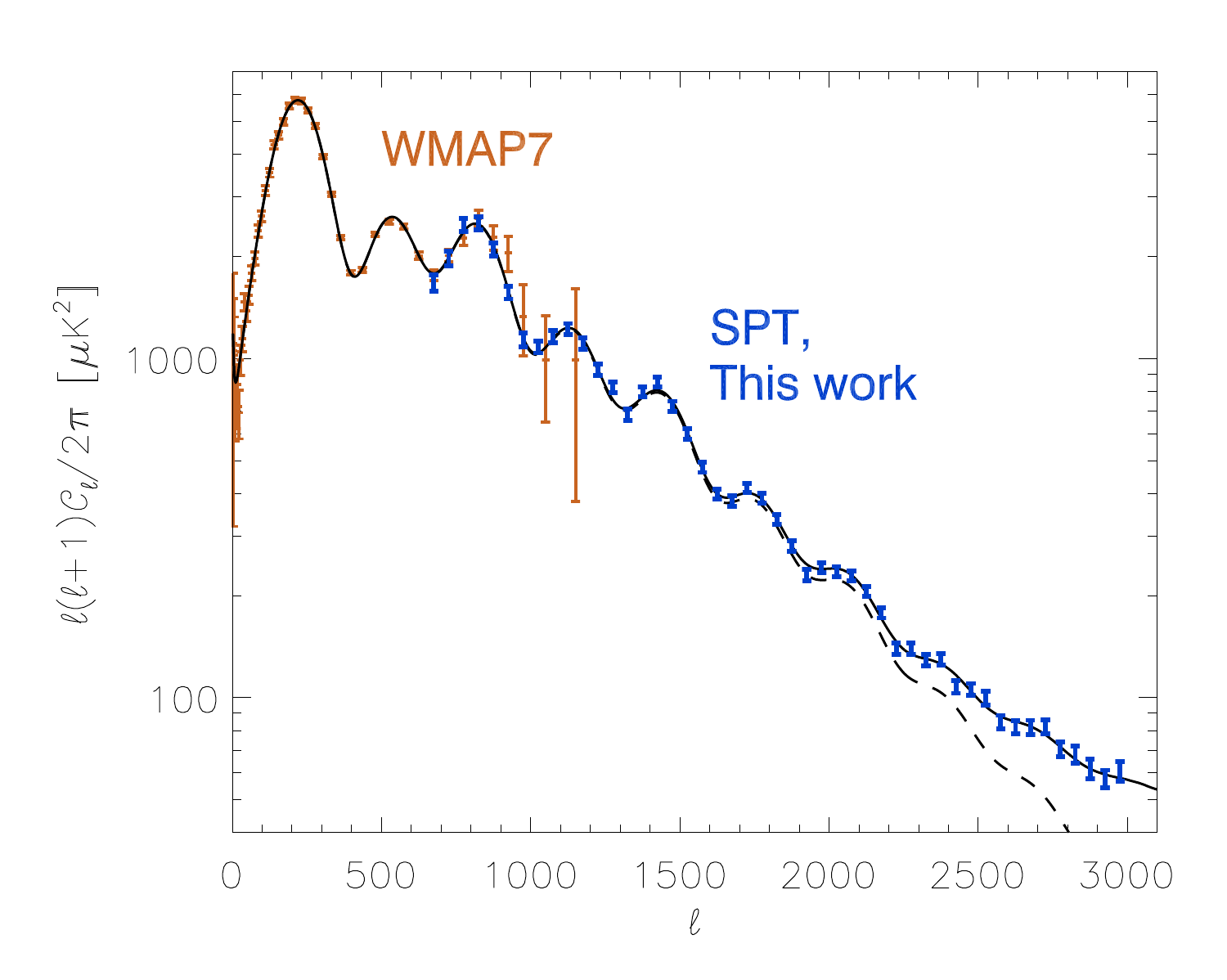} 
\includegraphics[width=80mm]{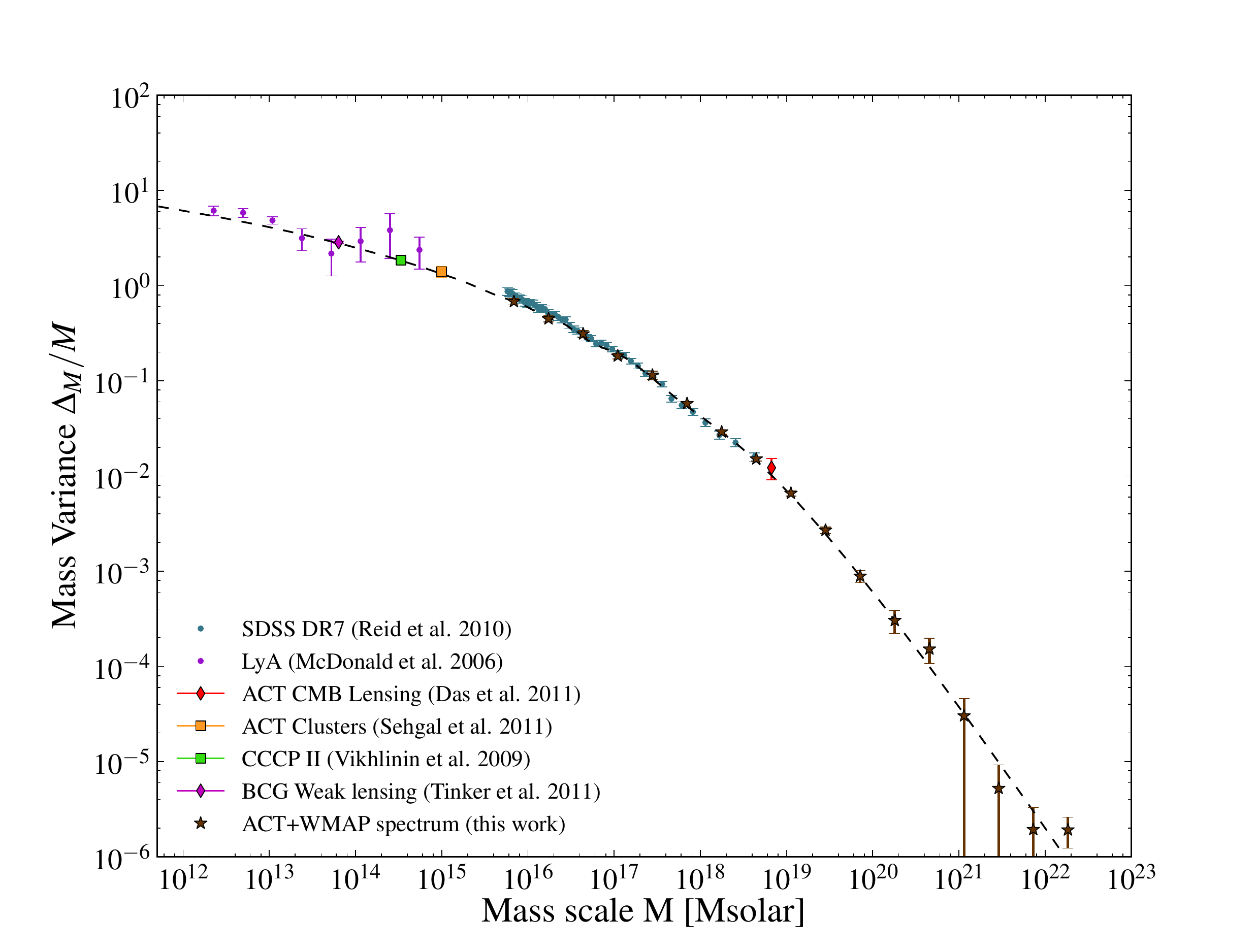} 
\caption{{\bf Left Panel:} Recent CMBR angular power spectrum from the WMAP 7-year data and from the South Pole Telescope 
observations show as band-averaged powers along with the best fit $\Lambda$CDM model (CMB - dashed line) and (CMB + 
foreground solid line), taken from Ref. \cite{Ke11}.  The detailed location, amplitude, and shape of the peaks (bumps) provide 
information on the contents of the Universe and the conditions at that early epoch plus secondary effects. 
{\bf Right panel:}  The mass variance of the reconstructed matter power spectrum from the Atacama Cosmology 
Telescope; large masses correspond to large scales and hence small values of $k$, taken from Ref. \cite{Hl11}.
The BAO are barely visible in the detail of this arching spectrum driven by the damping, mostly at low mass, of the 
original perturbations during the oscillations.}  
\label{power_spectra}
\end{figure}

\begin{theacknowledgments}
We gratefully acknowledge support from  CONACYT Grant No. 84133-F. 
\end{theacknowledgments}

\bibliographystyle{aipproc}   

\end{document}